\documentclass{llncs}
\usepackage{makeidx}  
\usepackage{graphicx}
\usepackage{amssymb,amsmath,amsfonts}
\begin{document}
%
%
%
%
%
\title{A Business Zone Recommender System Based on Facebook and Urban Planning Data}
%
%
\author{ Jovian Lin \and Richard~J.~Oentaryo  \and 
         Ee-Peng Lim \and Casey Vu \and \\Adrian Vu \and 
         Agus~T.~Kwee \and Philips~K.~Prasetyo          }
%

\tocauthor{Richard J. Oentaryo, Jia-Wei Low, and Ee-Peng Lim}
\institute{Living Analytics Research Centre, Singapore Management University\\
\email{jovian.lin@gmail.com, \{roentaryo, eplim, caseyanhthu, \\adrianvu, aguskwee, pprasetyo\}@smu.edu.sg}}

\maketitle              

\begin{abstract}
We present \texttt{ZoneRec}---a zone recommendation system for physical businesses in an urban city,
which uses both public business data from Facebook and urban planning data.
The system consists of machine learning algorithms that take in a business' metadata 
and outputs a list of recommended zones to establish the business in. 
We evaluate our system using data of food businesses in Singapore and 
assess the contribution of different feature groups to the recommendation quality.

\end{abstract}

\section{Introduction}


Location is a pivotal factor for retail success, owing to the fact that 94\% of retail sales are still transacted in physical stores \cite{Thau2015}. 
To increase the chance of success for their stores, business owners need to know not only where their potential customers are,
but also their surrounding competitors and potential allies.
However, assessing a store location is a cumbersome task for business owners as numerous factors need to be considered that often require gathering and analyzing the relevant data. 
To this end, business owners typically conduct ground surveys, which are time-consuming, costly, and do not scale up well. Moreover, with the rapidly changing environment and emergence of new business locations, one has to continuously reevaluate the value of the store locations. 

Fortunately, in the era of social media and mobile apps, we have an abundance of data that capture both online activities of users and offline activities at physical locations. 
For example, more than 890 million people actively use Facebook everyday \cite{Smith2014}. 
The availability of online user, location, and other behavioral data makes it possible now to estimate the value of a business location.

\begin{figure}[!t]
\centering
\includegraphics[width=0.825\textwidth]{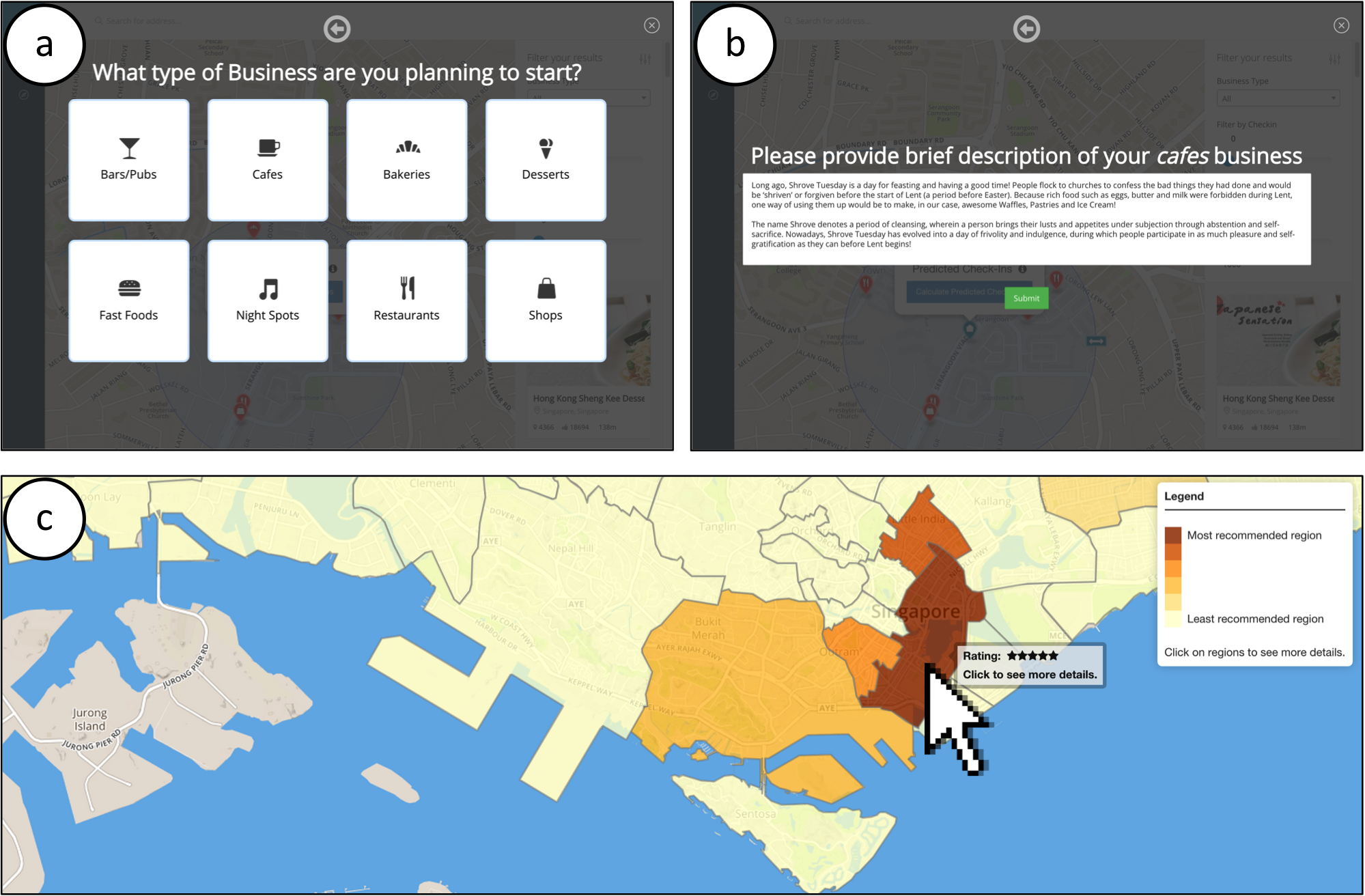}
\vspace{-5mm}
\caption{Our zone recommendation system.}
\label{figure:zone_recommendation_system}
\vspace{-6mm}
\end{figure}

Accordingly, we develop \texttt{ZoneRec}, a business location recommender system that takes a user's description about his/her business and produces a ranked list of \emph{zones} that would best suit the business.
Such ranking constitutes a fundamental information retrieval (IR) problem \cite{Manning2008,Liu2009}, where the user's description corresponds to the \emph{query}, and the pairs of business profiles and zones are the \emph{documents}. 
Our system is targeted at business owners who have little or no prior knowledge on which zone they should set up their business at.
In our current work, the zones refer to the 55 urban planning areas, the boundaries of which are set by the Singapore government. 
While we currently focus on Singapore data, it is worth noting that our approach can be readily used in other urban cities.

Fig.~\ref{figure:zone_recommendation_system} illustrates how our \texttt{ZoneRec} system works. First, the system asks the user to define the \emph{type} of his/her hypothetical (food) business (Fig.~\ref{figure:zone_recommendation_system}a),
and to then provide some \emph{description} of the business (Fig.~\ref{figure:zone_recommendation_system}b).
In turn, our system analyzes the input data, based on which its recommendation algorithm produces a ranked list of zones. 
The ranking scores of the zones are represented by a heatmap overlaid on the Singapore map (Fig.~\ref{figure:zone_recommendation_system}c). Further details of each recommended zone can be obtained by hovering or clicking on the zone.

\textbf{Related works}.
Using social media data to understand the dynamics of a society is an increasingly popular research theme.
For example, Chang and Sun~\cite{location3:2011} analyzed the ``check-ins'' data of Facebook users to develop models that can predict where users will check-in next, and in turn predict user friendships.
Another close work by Karamshuk \emph{et al.}~\cite{Karamshuk2013} demonstrated the power of geographic and user mobility 
features in predicting the best placement of retail stores. 
Our work differs from \cite{location3:2011} in that we use Facebook data to recommend locations instead of friendships.
Meanwhile, Karamshuk \emph{et al.} ~\cite{Karamshuk2013} discretized the city into a uniform grid of multiple circles. In contrast, we use more accurate, non-uniform area boundaries that are curated by government urban planning.

\textbf{Contributions}. In summary, our contributions are: 
(i) To our best knowledge, we are the first to develop a business zone recommendation method that fuses Facebook business location and urban planning data to help business owners find the optimal zone placement of their businesses; 
(ii) We develop a user-friendly web application to realize our \texttt{ZoneRec} approach, which is now available online at \texttt{http://research.larc.smu.edu.sg/bizanalytics/}. 
(iii) We conduct empirical studies to compare different algorithms for zone recommendation, and assess the relevancy of different feature groups.




\section{Datasets}
\label{sec:dataset}


In this work, we use two public data sources, which we elaborate below.

\textbf{Singapore urban planning data}. 
To obtain the zone information, we retrieved the urban planning data from the Urban Development Authority (URA) of Singapore \cite{URA2015}. 
The data consist of 55 predetermined planning zones.
To get the 55 zones, URA first divided Singapore into five regions: Central, West, North, North-East and East. Each region has a population of more than 500,000 people, and is a mix of residential, commercial, business and recreational areas. 
These regions are further divided into zones, each having a population of about 150,000 and being served by a town centre  and several  commercial/shopping centres.

\textbf{Facebook business data}.
In this work, we focus on data from Facebook pages about food-related businesses that are located within the physical boundaries of Singapore. 
Our motivation is that food-related businesses constitute one of the largest groups in our Singapore Facebook data.
From a total of 82,566 business profiles we extracted via Facebook's Graph API \cite{FBGraph2015}, we found 20,877 (25.2\%) profiles that are food-related.
Each profile has the following attributes:

\begin{itemize}
\item \textbf{Business name and description}. This represents the name and textual description of the shop, respectively.

\item \textbf{List of categories}.
From the 20,877 food-related businesses, we retrieve 357 unique categorical labels, as standardized by Facebook.
These may contain not only food-related labels such as ``bakery,'' ``bar,'' and ``coffee shop'',
but also non-food ones such as ``movie theatre,'' ``mall,'' and ``train station.''
The existence of non-food labels for food businesses is
Facebook's way of allowing the users to tag multiple labels for a business profile.
For example, a Starbucks outlet near a train station in an airport 
may have both food and non-food labels, 
such as ``airport,'' ``cafe,'' ``coffee shop,'' and ``train station''.
\item \textbf{Location of physical store}.
Each business profile has a \texttt{location} attribute containing
the physical address and latitude-longitude coordinates (hereafter called ``lat-long'').
We map the lat-long information to the URA data to determine which of the 55 zones the target business is in. 
Note that we rule out business profiles that do not have explicit lat-long coordinates.
%
\item \textbf{Customer check-ins}.
A check-in is the action of registering one's physical presence;
and the total number of check-ins received by a business gives us a rough estimate of how 
popular and well-received it is.
%
\end{itemize}

\section{Proposed Approach}

We cast the zone recommendation as a classification task, where the input features are derived from the textual and categorical information of a business and the class labels are the zone IDs. 
This formulation corresponds to the \emph{pointwise ranking} method for IR \cite{Liu2009}, whereby the ranking problem is transformed to a conventional classification task.
Our approach consists of three phases:

\textbf{Data cleaning}. 
      For each business, we first extract its (i) \emph{business name}, (ii) \emph{business description}, and (iii) the \emph{tagged categories} that it is associated with.
      As some business profiles may have few or no descriptive text, we set the minimum number of words in a description to be 20.
      This is to ensure that our study only includes quality business profiles, as the insertion of businesses with noisy ``check-ins'' will likely deterioriate the quality of the recommendations produced by our classification algorithms.
      We  remove all stop words and words containing digits.
      Stemming is also performed to reduce inflectional forms and derivationally related forms of a word to a common base form
      (\textit{e.g.}, car, cars, car's $\Rightarrow$ car). 
      
\textbf{Feature construction}.
      Using the cleaned text from the previous stage, we construct a bag of words for each feature group, \textit{i.e.}, the name, description, and categories of each business profile. 
      As not all words in the corpus are important, however,
      we compute the term frequency-inverse document frequency (TF-IDF) \cite{Manning2008} to measure how important a (set of) word or is to a business profile (\textit{i.e.}, a document) in a corpus.
      We also include bigram features, since in some cases pairs of words make more sense than the individual words.
      With the inclusion of unigrams and bigrams, we have a total number of 51,397 unique terms.
      We set the minimum document frequency (DF) as 3, 
      and retain the top 5000 terms based on their inverse document frequency (IDF) score.
      
\textbf{Classification algorithms}.
      Based on the constructed TF-IDF features of a business profile as well as the zone (\textit{i.e.}, class label) it belongs to, we can now craft the training data for our classification algorithms. Specifically, each classifier is trained to compute the \emph{matching score} between a business profile and a zone ID. We can then apply the classifiers to the testing data and sort the matching scores in descending order, based upon which we pick the $K$ highest scores that would constitute our top $K$ recommended zones.
      
      In this study, we investigate on three popular classification algorithms: 
      (i) support vector machine (SVM) with linear kernel (SVM-Linear) \cite{Chang2011}, 
      (ii) SVM with radial basis function kernel (SVM-RBF) \cite{Chang2011}, and
      (iii) random forest classifier (RF) \cite{Breiman2001}. 
      The first two aim at maximizing the margin of separation between data points from different classes, which would imply a lower generalization error. 
      Meanwhile, RF is an ensemble classifier that comprises an army of decision trees. It works based on bagging mechanism, \textit{i.e.}, each tree is built from bootstrap samples drawn with replacement from the training data, and the final prediction is done via a majority voting of the decisions made by the constituent trees.
      Being an ensemble model, RF exhibits its high accuracy and robustness, and the bagging mechanism facilitates an efficient, parallelizable learning process.

\section{Results and Analysis}

Our experiment aims at evaluating the quality of zone recommendation for new businesses. 
To do so, we hide some of the known businesses and assess the accuracy of the recommended zones for those businesses.
We measure the recommendation accuracy using three ranking metrics popularly used in IR, \textit{i.e.}, Hit@10, MAP@10, and NDCG@10 \cite{Manning2008}.
The Hit@10 calculates whether the actual zone ID is in the top 10 recommended zones, irregardless of the position of the actual zone ID. 
The MAP@10 and NDCG@10 compute the mean average precision and normalized discounted cumulative gain at top 10 respectively, both of which give higher penalty when the actual zone ID has a lower position in the recommendation list.
We evaluate our classifiers using 10-fold cross validation, whereby 90\% of the business profiles are used for training the  models, and the remaining 10\% for testing the models' performances on unseen profiles. We record the Hit@10, MAP@10, and NDCG@10 for each fold, and then report the averaged results.

\textbf{Performance assessment}. Table~\ref{comparison_against_baselines} shows the performances of different classifiers (with their corresponding best parameters).
Here RF substantially outperforms the two SVMs for all metrics, at a significance level of $0.01$ on the two-tailed $t$-test.
The superiority of RF over SVM can be explained by comparing ensemble vs. single models. 
First, by taking a consensus from different decision trees, RF reduces the risk of using a wrong classifier. In effect, the combined decision of multiple trees is more robust than that of a single tree. Also, the bagging mechanism helps reduce the modeling \emph{variance}---error from sensitivity to small fluctuations in the training data. Thus, RF is less prone to overfitting (\textit{i.e.}, modeling random noise in the data) than single models such as SVM.  

Comparing the two SVMs, we initially expected that SVM-RBF would outperform SVM-Linear, since the RBF kernel essentially maps the original data to an infinitely high-dimensional feature space, for which data from different classes should be more separable. 
It turns out, however, that SVM-Linear performs better than SVM-RBF.  
This may be attributed to our TF-IDF representation, which involves a sparse, fairly large number of features that is likely to be linearly separable already.
In such case, using nonlinear kernel would not necessarily help improve the performance, and may instead increase the risk of overfitting.

\begin{table}[!t]
\scriptsize
\centering
\caption{Recommendation results of different algorithms}
\label{comparison_against_baselines}
\begin{tabular}{|l|c|c|c|}
\hline
\textbf{Algorithm} & \textbf{~Hit@10~} & \textbf{~MAP@10~} & \textbf{~NDCG@10~} \\ 
\hline
SVM-Linear ($C=1$)         & 0.502              & 0.300              & 0.348               \\
SVM-RBF ($C=1$, $\gamma=\frac{1}{\text{\#features}}$)           & 0.231              & 0.074              & 0.110               \\
Random forest ($\text{\#trees}=1000$)       & \textbf{0.721*}   & \textbf{0.430*}   & \textbf{0.499*}    \\
\hline
\multicolumn{4}{l}{$C$: cost parameter, $\gamma$: kernel coefficient, *: significant at $0.01$}
\end{tabular}
\end{table}

\textbf{Contribution of features}. As mentioned in Section \ref{sec:dataset}, we divide our feature vectors into three groups: (i) business name, (ii) business description, and (iii) tagged categories. Here we evaluate the contribution of each feature group by performing an \emph{ablation test} on the RF model. Table~\ref{figure:ablate} consolidates the results of our ablation study. The first three rows show the results of ablating (removing) two feature groups, while the last three rows are for ablating one feature group. 

\begin{table}[!t]
\scriptsize
\centering
\caption{Feature ablation results for random forest classifier}
\label{figure:ablate}
\begin{tabular}{|c|c|c|c|c|c|}
\hline
\textbf{Use Name}  & \textbf{Use Description} & \textbf{Use Categories} & \textbf{Hit@10} & \textbf{MAP@10} & \textbf{NDCG@10} \\
\hline
-  & -  & Yes  & 0.537        & 0.212        & 0.287         \\
-  & Yes  & -  & 0.685  & 0.398        & 0.465         \\ 
Yes  & -  & -  & 0.554        & 0.240        & 0.313          \\
\hline
Yes  & -   & Yes    & 0.557        & 0.239        & 0.313         \\
Yes  & Yes   & -    & \textbf{0.708}        & \textbf{0.423}        & \textbf{0.499}         \\
-  & Yes   & Yes    & 0.694        & 0.404        & 0.472          \\
\hline
\end{tabular}
\vspace{-3mm}
\end{table}

From the first three rows, it is evident that the ``description'' is the most important feature group, consistently providing the highest Hit@10, MAP@10, and NDCG@10 scores compared to the other two.
This is reasonable, as the ``description'' provides the richest set of features (in terms of word vocabulary and frequencies) representing a business, and some of these features provide highly discriminative inputs for our RF classifier.
We can also see that the ``name'' group is more discriminative than the ``categories'' group for all the three metrics. 
Again, this can be attributed to the more fine-grained information provided by the business' name features as compared to the category features.
Finally, we find that the results in the last three rows are consistent with those of the first three rows. That is, the ``description'' group constitutes the most informative features (for our RF model), followed  by the ``name'' and ``category'' groups.

\section{Conclusion}

We put forward the \texttt{ZoneRec} recommender system that can help business owners decide which zones they should set their businesses at.
Despite its promising potentials, there remains room for improvement. First, the zone-level recommendations may not provide sufficiently granular information for business owners, \textit{e.g.}, where exactly a store should be set at and how the surrounding businesses may affect this choice. 
It is also fruitful to include more comprehensive residential and demographic information in our feature set, and conduct deeper analysis on the contribution of the individual features.
To address these, we plan to develop a two-level location recommender system, whereby \texttt{ZoneRec} serves as the first level and the second level recommends the specific hotspots within each zone. 


\subsubsection{Acknowledgments.} \small This research is supported by the Singapore National Research Foundation under its International Research Centre @ Singapore Funding Initiative and administered by the IDM Programme Office, Media Development Authority (MDA).

\bibliographystyle{abbrv}
\bibliography{llncs}
\end{document}